\documentclass[twocolumn]{aastex61}

\shorttitle{Modeling of the Kilonova Associated with GW170817}
\shortauthors{Villar et al.}
\usepackage{hyperref}
\usepackage{cleveref}
\usepackage{graphicx}
\usepackage{amsmath}
\usepackage{amssymb}
\usepackage{pbox}
\usepackage{color}
\usepackage{xspace}
\usepackage{apjfonts} 
\usepackage{array}

\hypersetup{colorlinks,citecolor=blue}

\newcommand\mosfit{{\tt MOSFiT}\xspace}
\newcolumntype{C}[1]{>{\centering\let\newline\\\arraybackslash\hspace{0pt}}m{#1}}
\newcolumntype{T}[1]{>{\centering\let\newline\\\arraybackslash\hspace{0pt}}p{#1}}
\newcolumntype{P}[1]{>{\raggedright\arraybackslash}p{#1}}

\begin{document}

\title{The Combined Ultraviolet, Optical, and Near-Infrared Light Curves of the Kilonova Associated with the Binary Neutron Star Merger GW170817:  Unified Data Set, Analytic Models, and Physical Implications}

\author{V.~A.~Villar}
\affiliation{Harvard-Smithsonian Center for Astrophysics, 60 Garden Street, Cambridge, Massachusetts 02138, USA}
\author{J.~Guillochon}
\affiliation{Harvard-Smithsonian Center for Astrophysics, 60 Garden Street, Cambridge, Massachusetts 02138, USA}
\author{E.~Berger}
\affiliation{Harvard-Smithsonian Center for Astrophysics, 60 Garden Street, Cambridge, Massachusetts 02138, USA}
\author{B.~D.~Metzger}
\affiliation{Department of Physics and Columbia Astrophysics Laboratory, Columbia University, New York, NY 10027, USA}
\author{P.~S.~Cowperthwaite}
\affiliation{Harvard-Smithsonian Center for Astrophysics, 60 Garden Street, Cambridge, Massachusetts 02138, USA}
\author{M.~Nicholl}
\affiliation{Harvard-Smithsonian Center for Astrophysics, 60 Garden Street, Cambridge, Massachusetts 02138, USA}
\author{K.~D.~Alexander}
\affiliation{Harvard-Smithsonian Center for Astrophysics, 60 Garden Street, Cambridge, Massachusetts 02138, USA}
\author{P.~K.~Blanchard}
\affiliation{Harvard-Smithsonian Center for Astrophysics, 60 Garden Street, Cambridge, Massachusetts 02138, USA}
\author{R.~Chornock}
\affiliation{Astrophysical Institute, Department of Physics and Astronomy, 251B Clippinger Lab, Ohio University, Athens, OH 45701, USA}
\author{T.~Eftekhari}
\affiliation{Harvard-Smithsonian Center for Astrophysics, 60 Garden Street, Cambridge, Massachusetts 02138, USA}
\author{W.~Fong}
\altaffiliation{Hubble Fellow}
\affiliation{Center for Interdisciplinary Exploration and Research in Astrophysics (CIERA) and Department of Physics and Astronomy, Northwestern University, Evanston, IL 60208}
\author{R.~Margutti}
\affiliation{CIERA and Department of Physics and Astronomy, Northwestern University, Evanston, IL 60208}
\author{P.~K.~G.~Williams}
\affiliation{Harvard-Smithsonian Center for Astrophysics, 60 Garden Street, Cambridge, Massachusetts 02138, USA}

\begin{abstract}
We present the first effort to aggregate, homogenize, and uniformly model the combined ultraviolet, optical, and near-infrared dataset for the electromagnetic counterpart of the binary neutron star merger GW170817. By assembling all of the available data from 18 different papers and 46 different instruments, we are able to identify and mitigate  systematic offsets between individual datasets, and to identify clear outlying measurements, with the resulting pruned and adjusted dataset offering an opportunity to expand the study of the kilonova.  The unified dataset includes 647 individual flux measurements, spanning $0.45$ to $29.4$ days post-merger, and thus has greater constraining power for physical models than any single dataset. We test a number of semi-analytical models and find that the data are well modeled with a three-component kilonova model: a ``blue'' lanthanide-poor component ($\kappa=0.5$ cm$^2$ g$^{-1}$) with $M_\mathrm{ej}\approx 0.020$ M$_\odot$ and $v_\mathrm{ej}\approx 0.27 c$; an intermediate opacity ``purple'' component ($\kappa=3$ cm$^2$ g$^{-1}$) with $M_\mathrm{ej}\approx 0.047$ M$_\odot$ and $v_\mathrm{ej}\approx 0.15 c$; and a ``red'' lanthanide-rich component ($\kappa=10$ cm$^2$ g$^{-1}$) with $M_\mathrm{ej}\approx 0.011$ M$_\odot$ and $v_\mathrm{ej}\approx 0.14 c$. We further explore the possibility of ejecta asymmetry and its impact on the estimated parameters. From the inferred parameters we draw conclusions about the physical mechanisms responsible for the various ejecta components, the properties of the neutron stars, and, combined with an up-to-date merger rate, the implications for $r$-process enrichment via this channel. To facilitate future studies of this keystone event we make the unified dataset and our modeling code public.
\end{abstract}

\section{Introduction}
\label{sec:intro}

The joint detection of gravitational waves and electromagnetic radiation from the binary neutron star merger GW170817 marks the beginning of a new era in observational astrophysics. The merger was detected and localized by the Advanced LIGO and Virgo detectors to a sky region of about 30 deg$^2$ at a distance of $\approx 24-48$ Mpc, with inferred component masses of $\approx 1.36-1.60$ and $\approx 1.17-1.36$ M$_\odot$ (90\% confidence ranges for the prior of low neutron star spins; \citealt{abbott2017gw170817}). A spatially coincident short-duration gamma-ray burst (SGRB) was detected with a delay of 1.7 seconds relative to the merger time \citep{LIGOGRBPaper, goldstein2017ordinary, IntegralDetection}. About 11 hours post-merger several groups \citep{abbott2017multi,SwopeDiscovery,soares2017electromagnetic,DLT40Discovery} independently detected an optical counterpart coincident with the quiescent galaxy NGC\,4993 at a distance of $39.5$ Mpc \citep{freedman2001final}. 

Subsequently, multiple ground- and space-based observatories followed up the optical counterpart in the UV, optical, and NIR (hereafter, UVOIR), extending to about $30$ days post-merger when the location of the source near the Sun prevented further observations.  These observations were published in multiple papers that appeared when the detection was publicly announced on October 16, 2017 \citep{andreoni2017,arcavi2017optical,SwopeDiscovery,cowperthwaite2017electromagnetic,diaz2017observations,drout2017light,evans2017swift,hu2017optical,kasliwal2017illuminating,lipunov2017master,pian2017spectroscopic,pozanenko2017grb170817a,ShappeeSpectra,smartt2017kilonova,tanvir2017emergence,troja2017x,utsumi2017j,valenti2017discovery}.  The various papers generally conclude that the UVOIR emission is due at least in part to a kilonova, a quasi-thermal transient powered by the radioactive decay of newly-synthesized $r$-process nuclei and isotopes \citep{li1998transient,metzger2010electromagnetic,roberts2011electromagnetic,metzger2012most,barnes2013effect,tanaka2013radiative}. In particular, there is general agreement that the observed light curves require at least two distinct components: a ``blue'' component that dominates the emission in the first few days, followed by a transition to a ``red'' component. This multi-component behavior is also seen in optical and NIR spectroscopic observations of the transient \citep{chornock2017electromagnetic,nicholl2017electromagnetic,pian2017spectroscopic,ShappeeSpectra,smartt2017kilonova}. The blue emission is interpreted to be due to ejecta dominated by Fe-group and light $r$-process nuclei (atomic mass number $A\lesssim 140$), while the red emission is likely due to ejecta rich in lanthanides and heavy $r$-process material ($A\gtrsim 140$).

In \citet{cowperthwaite2017electromagnetic}, we modeled photometric data from the Dark Energy Camera (DECam), {\it Swift}/UVOT, Gemini, and the {\it Hubble Space Telescope} ({\it HST}) using the flexible light curve modeling code \mosfit \citep{GuillochonMosfit}. The analysis demonstrated that the UVOIR data cannot be explained by the radioactive decay of $^{56}$Ni, nor with the associated opacity from Fe-peak elements alone. The data could be well matched by a kilonova model using $r$-process heating but required at least two distinct components (red and blue) with different opacities, masses, and velocities. A model with a third component (with a higher lanthanide fraction) fit the data equally well \citep{cowperthwaite2017electromagnetic}. A similar conclusion was reached by several other groups modeling independent sets of observations (e.g., \citealt{2017arXiv171005850T,2017arXiv171005434K}). However, given our limited dataset, we were unable to break degeneracies between the two- and three-component models.

Following the publication of multiple datasets, we undertake here the first effort to aggregate, homogenize, and model all of the available UVOIR measurements. In total, the UVOIR dataset includes $714$ individual measurements from $46$ different instruments.  After collecting the data, we identify measurements that are clearly discrepant from the majority of similar observations, and where possible correct for systematic deviations in order to include as many photometric points as possible. The final unified dataset includes $647$ measurements. With this extensive dataset we revisit the models first explored in \citet{cowperthwaite2017electromagnetic} with a number of refinements to the physical setup; the model setup is available via the Open Kilonova Catalog\footnote{\url{https://kilonova.space/} \citep{OSCRef}.} (OKC).

The layout of the paper is as follows: In Section~\ref{sec:data} we discuss the various datasets and describe our approach to standardize the data.  In Section~\ref{sec:model} we present our model, including additional parameters designed to capture possible asymmetries in the ejecta geometry.  We present the results of the model fits in Section~\ref{sec:results} and explore their implications in Section~\ref{sec:implications}.

\begin{deluxetable*}{P{2.7cm}P{2.6cm}P{3.0cm}P{2.6cm}P{2.6cm}P{3.0cm}} 
\tabletypesize{\footnotesize} 
\tablecolumns{6} 
\tablewidth{0pt} 
\tablecaption{Data Summary \label{table:data_words}} 
\tablehead{ \colhead{Reference} &
\colhead{Bands} & 
\colhead{Instruments} &
\colhead{Telescopes} &
\colhead{Photometry} &
\colhead{Comments} \\ }
\startdata 
\citeauthor{andreoni2017} & $g$,$r$,$i$,C & SkyMapper, 2k2k CCD, 1k2k CCD, NAOS-CONICA, VISIR & SkyMapper, Zadko, VIRT, VLT & image subtraction & 
  Additional data to be published by authors.\\[0.001cm]
\citeauthor{arcavi2017optical} & $V$, $g$, $r$, $i$, $z$, $w$ & Sinistro & LCO 1m/CTIO, SAAO, Siding Spring & image subtraction & 
  Possible template contamination in $V$-, $g$-, $r$-, and $i$-band; $w$-band calibrated using $r$-band SDSS reference stars\\[0.001cm]
\citeauthor{SwopeDiscovery} & $B$, $V$, $g$, $r$, $i$ & E2V 4k4k CCD & Swope & PSF-fitting & \\[0.001cm]
\citeauthor{cowperthwaite2017electromagnetic} & $u$, $g$, $r$, $i$, $z$, $Y$ & DECam& Blanco/CTIO, & image subtraction & \\[0.001cm]
\citeauthor{cowperthwaite2017electromagnetic} & $F336W$, $F475W$, $F625W$, $F775W$, $F850LP$, $F110W$, $F160W$, $H$, $K_\mathrm{s}$ & WFC3/UVIS, ACS/WFC, WFC3/IR, Flamingos-2  & \textit{HST}, Gemini-South  & PSF-fitting & \\[0.001cm]
\citeauthor{diaz2017observations} & $g$, $r$, $i$ & T80Cam & T80S/CTIO & PSF-fitting & \\[0.001cm]
\citeauthor{drout2017light} & $B$, $g$, $r$, $i$, $z$, $J1$, $J$, $H$, $K_\mathrm{s}$ & IMACS, LDSS-3, FourStar, RetroCam & Magellan,  du Pont & PSF-fitting & Used rotated image of galaxy as template \\[0.001cm]
\citeauthor{drout2017light} & $U$, $V$, $g$, $I$, $J$, $H$, $K_\mathrm{s}$ & EFOSC2, SOFI, LRIS & NTT, Keck-I & PSF-fitting & \\[0.001cm]
\citeauthor{evans2017swift} & $UVW2$, $UVM2$, $U$, $B$, $V$ & UVOT & Swift & host count rate subtraction & \\[0.001cm]
\citeauthor{hu2017optical} & i & 10k10k CCD & AST3-2 & image subtraction & Possible template contamination in $i$-band\\[0.001cm]
\citeauthor{valenti2017discovery} & $r$ & Alta U47+ & PROMPT5 & image subtraction & Pre-existing template \\[0.001cm]
\citeauthor{kasliwal2017illuminating} & $F225W$, $F336W$, $B$, $g$, $V$, $r$, $R$, $i$, $I$, $z$, $u$, $J$, $H$, $K_\mathrm{s}$ & Flamingos-2, GMOS, WIRC, SIRIUS, ANDICAM, NICFPS, VISIR, WFC3/UVIS & Gemini, Palomar, IRSF, CTIO 1.3m, APO 3.5m, VLT, HST  & PSF-fitting, aperture photometry & Subtraction of median-filtered image to remove galaxy \\[0.001cm]
\citeauthor{lipunov2017master} & $B, V, R, W$ & MASTER & OAFA, SAAO &image subtraction & Pre-existing template\\[0.001cm]
\citeauthor{pian2017spectroscopic} & $B, V, g, r, R, i, I, z$ & FORS2, ROS2, X-shooter, OmegaCam & VLT, VST, REM & PSF-fitting & \\[0.001cm]
\citeauthor{pozanenko2017grb170817a}& \tt{LUM} & 4k4k CCD & RC-1000 & image subtraction & {\tt LUM}-band calibrated using $r$-band reference stars \\[0.001cm]
\citeauthor{ShappeeSpectra} & $B, V, R, I$ , $g, r, i, z$ & IMACS, LDSS-3 & Magellan & synthetic photometry & Generated synthetic photometry from spectra \\[0.001cm]
\citeauthor{smartt2017kilonova}& $g, r, i, z, y, J, H, K$ & GFC, EFOSC2 & Pan-STARRS, NTT, 1.5B & image subtraction & Pre-existing template  \\[0.001cm]
\citeauthor{smartt2017kilonova} & $U, g, r, i, z, J, H, K$ & GROND & MPI/ESO 2.2m & image subtraction & Possible template contamination in GROND $K$-band\\[0.001cm]
\citeauthor{tanvir2017emergence} & F475W, $r$, F606W, $i$, F814W, $z, Y, J$, F110W, F160W, $K_\mathrm{s}$ & VIMOS, WFC-UVIS, FORS, DK1.5, VISTA, NOTCam, WFC-IR, HAWK-I & HST, VLT, HST, DK1, VISTA, NOT & aperture photometry & Local background subtraction; F110W calibrated to $J$-band. \\[0.001cm]
\citeauthor{troja2017x} & F275W, $B, V$, F475W, F606W, $R$, $I$, $z$, $J$, $H$, $K_\mathrm{s}$, F110W, F160W & WFC-IR, WFC-UVIS, GMOS & HST, KMTNet, Gemini & image subtraction & \\[0.001cm]
\citeauthor{utsumi2017j} & $V, R, g, r, i, z, J, H, K$ & HSC, SIRIUS, MOA-II, MOACam, MOIRCS & B\&C, IRSF, Tripol5, Subaru & PSF-fitting & MOACam $R$-band converted to standard $R$-band using empirical relationship \\
\enddata 
\end{deluxetable*}

\section{Ultraviolet, Optical, and Near-Infrared Data}
\label{sec:data}

Following the public announcement of the discovery and observations of GW170817, we aggregated the UVOIR photometry available in the literature, which we provide in this paper and in the OKC. The data span from $0.45$ days to $29.4$ days post-merger, and were collected with $46$ instruments in $37$ unique filters. This extensive dataset represents a departure from most transient light curves, with over twenty observations taken each night on average with fairly complete color coverage during the duration of the event. For each published set of observations, we summarize the instruments and filters used, the details of the photometry methods, and any relevant notes in Table~\ref{table:data_words}. All photometry is reported as AB magnitudes with no correction for Milky Way extinction.

Thanks to the extensive observations from multiple telescopes there is significant redundancy of photometric measurements.  This allows us to compare individual datasets to the bulk of the other observations and hence to homogenize and prune the dataset.  With this approach we find that some corrections are required for three datasets: $gri$-band data from \citet{arcavi2017optical}, some $K_\mathrm{s}$-band data from \citealt{smartt2017kilonova} and $i$-band data from \citealt{hu2017optical}. All of these datasets utilized image subtraction to isolate the flux of the transient.  However, we find that for the specific filters listed above the resulting light curves were typically dimmer, and faded more rapidly, than the rest of the data.  We interpret this as being due to residual emission from the transient in the reference templates, since in each case the template was obtained after the discovery of the source (however it is also possible that the PSF photometry is contaminated by residual host flux). Using the dates of the template images (Arcavi, private communication, \citealt{smartt2017kilonova} and \citealt{hu2017optical}), we estimate the kilonova brightness for each filter and add this residual flux to the reported photometry.  Specifically, we use estimated template magnitudes of: 20.8 ($g$), 20.9 ($r$), 20.3 ($i$) and 20.0 ($z$) mag to the \citet{arcavi2017optical} dataset; 19.4 ($K_s$, GROND data only) mag to the \citet{smartt2017kilonova} dataset; and 19.9 ($i$) mag to the \cite{hu2017optical} dataset.  With these corrections the data are in good agreement with the photometry from other sources (to $\lesssim 0.2$ mag). With better template images, the residual systematic differences should diminish.

We additionally exclude two datasets from our model fitting: the $r$-band dataset from \citet{pozanenko2017grb170817a}, which was obtained in the {\tt LUM} filter but calibrated to $r$-band reference stars; and the $w$-band from \citet{arcavi2017optical}, which was similarly calibrated using $r$-band reference stars. Because the kilonova colors differ so drastically from the comparison stars (see e.g., \citealt{cowperthwaite2017electromagnetic}), these calibrations are  unreliable. 

Due to the fact that the observations conducted by the {\it Swift} UV/Optical Telescope (UVOT) were publicly available, three papers presented independent analyses and photometry of these data \citep{cowperthwaite2017electromagnetic,drout2017light,evans2017swift}.  However, in our homogenized dataset we only use the photometry presented by the {\it Swift} team \citep{evans2017swift} without alteration. Early photometry is largely consistent among the three papers to within $\approx 0.2$ mag, although the reported observation times differ by several hours due to different choices of time binning. 

Similarly, several teams independently analyzed some Gemini-South FLAMINGOS-2 data \citep{cowperthwaite2017electromagnetic,kasliwal2017illuminating,troja2017x}, some NTT EFOSC2 data \citep{drout2017light,smartt2017kilonova}, and some {\it HST}/WFC3 data \citep{tanvir2017emergence,troja2017x}. All of the measurements are listed in Table~\ref{table:data} but marked as repeated observations.  The {\it HST}/WFC3/F110W data from \cite{tanvir2017emergence} are re-calibrated to ground-based $J$-band photometry, so we use the data for these epochs from \cite{troja2017x}.  For all other epochs with multiple analyses of the same data we take a weighted average of the reported photometry for use in the model fitting, excluding outliers (see below); we  report the averaged values in Table~\ref{table:data}. 

Finally, we identify individual outlying data points through visual inspection and comparison. In total, we find fifteen such data points. Three of these are photometry of common data analyzed by multiple teams, so we simply exclude these points from our averaged photometry. We include the twelve other outliers in our modeling, but specifically identify these outliers in Table~\ref{table:data}.

The combined dataset is listed in Table~\ref{table:data}.  This table includes the MJD date and phase of each observation; the instrument, telescope, and filter combination; our corrected magnitudes and uncertainties; the correction applied to the original magnitudes (where applicable); a reference to the original paper; and a note indicating if the data were excluded from modeling (``X''), were included in modeling (``*''), represent a repeated reduction of the same observations (``R''), are averaged values from repeated observations (``A''), or are marked as outliers (``O''). {\it We request that any use of the data in this table includes appropriate citation to the original papers, as well as to our compilation}.

To properly model this extensive and heterogeneous dataset we use the appropriate transmission curve (or close equivalent) for each filter, instrument, and telescope combination\footnote{All transmission curves used in this work were obtained through the Spanish Virtual Observatory, \url{http://svo2.cab.inta-csic.es/svo/theory/fps3/} \citep{svo2012}, which aggregates official transmission curves for each instrument.}.

Photometric modeling of the host galaxy, NGC\,4993, suggests that the host environment contributes minimal extinction \citep{blanchard2017electromagnetic}\footnote{\citet{levan2017} find evidence for more moderate extinction, $E(B-V)=0.07$ mag, from spectroscopic observations near the explosion site.}. We therefore only include a correction for Milky Way extinction, with $E(B-V)=0.105$ mag \citep{schlafly2011measuring}.

\section{Kilonova Model}
\label{sec:model}

In this section we outline the analytical kilonova model first introduced in \citet{metzger2016kilonova} and implemented in \mosfit by \citet{villar2017}.  This model was also used in \citet{cowperthwaite2017electromagnetic} to model our own set of observations.

Following decompression from high densities, seed nuclei within the neutron-rich ejecta from a BNS merger undergo rapid neutron capture ($r$-process) nucleosynthesis \citep{li1998transient,metzger2010electromagnetic}, and it is the radioactive decay of these freshly-synthesized nuclei that powers the kilonova \citep{metzger2016kilonova}.  Unlike SNe, which are powered primarily by the radioactive decay of one species ($^{56}$Ni) and therefore undergo exponential decline in their bolometric light curves, kilonovae are powered by the decay of a wide range of $r$-process nuclei with different half-lives, leading to a power-law decay.  At very early times (first few seconds), the energy generation rate is roughly constant as neutrons are consumed during the $r$-process, but subsequently the $r$-process freezes out and the energy generation rate approaches a power-law decay, $\propto t^{-\alpha}$ with $\alpha\approx 1.3$ \citep{metzger2010electromagnetic}.  The temporal evolution of the radioactive heating rate can be approximated by the parameterized form \citep{korobkin2012astrophysical}: 
\begin{multline}
    L_\mathrm{in}(t) = 4\times10^{18}M_\mathrm{\rm rp}\times\\
    \left[0.5 - \pi^{-1}\arctan\left(\frac{t-t_0}{\sigma}\right)\right]^{1.3} \text{erg s}^{-1},
\end{multline}
where $M_\mathrm{\rm rp}$ is the mass of the $r$-process ejecta, and $t_0=1.3$ s and $\sigma=0.11$ s are constants. Our chosen input luminosity described above neglects any contribution from fall-back accretion on the newly formed remnant. Hydrodynamical simulations suggest that disk winds prevent the fall-back material from reaching the remnant on timescales $\gtrsim100$ ms \citep{Fernandez&Metzger13,metzger2016kilonova}; however, some contribution to the bolometric light curve from fall-back accretion is possible on longer (days to weeks) timescales.

Although $L_{\rm in}$ provides the total power of radioactive decay (shared between energetic leptons, $\gamma$-rays, and neutrinos), only a fraction $\epsilon_{\rm th}<1$ of this energy thermalizes within the plasma and is available to power the kilonova \citep{metzger2010electromagnetic}.  The thermalization efficiency decreases as the ejecta become more dilute with time, in a manner that can be approximated analytically as \citep{barnes2016radioactivity}:
\begin{equation}
    \epsilon_{\rm th}(t) = 0.36\left[e^{-at}+\frac{\ln(1+2bt^{d})}{2bt^{d}}\right],
\end{equation}
where $a$, $b$, and $d$ are constants of order unity that depend on the ejecta velocity and mass. We use an interpolation of Table 1 of \cite{barnes2016radioactivity} for these values.  

Assuming that the energy deposition is centrally located and the expansion is homologous, we can use the formalism originally outlined in \cite{arnett1982type} to compute the observed bolometric luminosity \citep{chatzopoulos2012generalized}:
\vspace{-0.0cm}
\begin{equation}
 L_{\rm bol}(t) = \exp\left(\frac{-t^2}{t_d^2}\right)\times
 \int_0^t L_{\rm in}(t) \epsilon_{\rm th}(t) \exp\left(t^2/t_d^2\right)\frac{t}{t_d} {\rm d}t,
\end{equation}
where $t_\mathrm{d}\equiv\sqrt{2\kappa M_\mathrm{rp}/\beta vc}$, $\kappa$ is the grey opacity, and $\beta=13.4$ is a dimensionless constant related to the ejecta mass geometric profile. We note that the assumption of a centrally concentrated power source is not necessarily true for kilonovae, as here we assume that the ejecta consists entirely of radioactive $r$-process material. Relaxation of this assumption should be explored in future work.

We explore multi-component models in which each component has a different opacity corresponding to theoretical expectations for different ejecta compositions. The opacity is largely determined by the fraction of lanthanides in the ejecta, with lanthanide-poor ejecta having a typical opacity of $\kappa\approx 0.5$ cm$^2$ g$^{-1}$, and lanthanide-rich ejecta having a typical opacity of $\kappa\approx 10$ cm$^2$ g$^{-1}$ \citep{tanaka2017properties}. A larger opacity results in a slower light curve evolution and a shift of the spectral energy distribution peak to redder wavelengths. We specifically explore a model with two components (``blue'', $\kappa=0.5$ cm$^2$ g$^{-1}$ and ``red'', $\kappa$ left as a free parameter), and with three components (``blue'', $\kappa=0.5$ cm$^2$ g$^{-1}$; ``purple'', $\kappa=3$ cm$^2$ g$^{-1}$ and ``red'', $\kappa=10$ cm$^2$ g$^{-1}$; \citealt{tanaka2017properties}).  The purple component corresponds to ejecta with a low, but non-negligible, lanthanide fraction. Each component of the multi-component model is evolved independently, accounting for the unique opacities and therefore diffusion timescales.

To model the multi-band light curves, we assume that each component has a blackbody photosphere with a radius that expands at a constant velocity ($v_\mathrm{phot}\equiv v$, where $v$ is the ejecta velocity).  At every point in time, the temperature of each component is defined by its bolometric luminosity and radius, using the Stefan-Boltzmann law. However, when the ejecta cool to a critical temperature ($T_c$) the photosphere recedes into the ejecta and the temperature remains fixed. The full SED of the transient is given by the sum of the blackbodies representing each component. The blackbody approximation and temperature floor behavior have both been seen in more sophisticated simulations \citep{barnes2013effect}; the temperature floor may relate to the first ionization temperature in lanthanide species. The analytic form of the blackbody behavior is: 
\begin{equation}
T_\mathrm{phot}(t) = \max\Big[\left(\frac{L(t)}{4\pi\sigma^2_\mathrm{SB}v_\mathrm{ej}^2t^2}\right)^{1/4},T_c\Big],
\end{equation}
and
\begin{equation}
R_\mathrm{phot}(t) = \begin{cases}
v_\mathrm{ej}t &\left(\frac{L(t)}{4\pi\sigma^2_\mathrm{SB}v_\mathrm{ej}^2t^2}\right)^{1/4}>T_c\\
\left(\frac{L(t)}{4\pi\sigma_\mathrm{SB}T_\mathrm{c}^4}\right)^{1/2} &\left(\frac{L(t)}{4\pi\sigma^2_\mathrm{SB}v_\mathrm{ej}^2t^2}\right)^{1/4}\leq T_c
\end{cases}
\end{equation}

\begin{figure*}[t]
\centering
\includegraphics[width=0.99\textwidth]{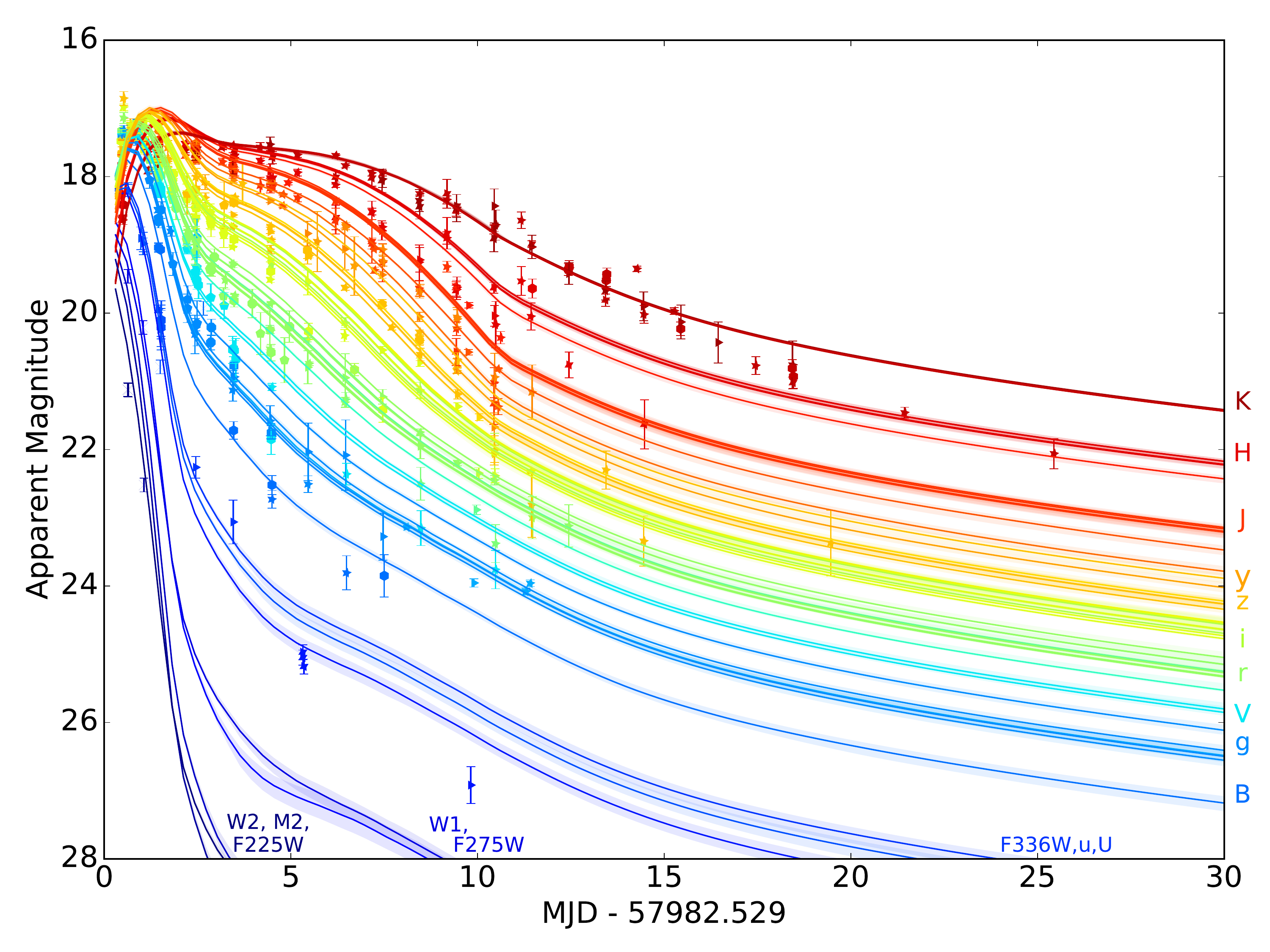}
\caption{UVOIR light curves from the combined dataset (Table~\ref{table:data}), along with the spherically symmetric three-component models with the highest likelihood scores. Solid lines represent the realizations of highest likelihood for each filter, while shaded regions represent the $1\sigma$ uncertainty ranges. For some bands there are multiple lines that capture subtle differences between filters. Data originally presented in \citealt{andreoni2017,arcavi2017optical,SwopeDiscovery,cowperthwaite2017electromagnetic,diaz2017observations,drout2017light,evans2017swift,hu2017optical,kasliwal2017illuminating,lipunov2017master,pian2017spectroscopic,pozanenko2017grb170817a,ShappeeSpectra,smartt2017kilonova,tanvir2017emergence,troja2017x,utsumi2017j,valenti2017discovery}.}\label{fig:lcs}
\end{figure*}

\subsection{Asymmetric Model}

In addition to the spherically symmetric assumption in the previous section we also explore a simple asymmetric model in which the blue component is confined to the polar regions, while the red component (and purple component in the three-component model) are confined to an equatorial torus. Such a model is seen in numerical simulations (see e.g., \citealt{Metzger&Fernandez14,metzger2016kilonova}).  We implement this asymmetric distribution by correcting the bolometric flux of each component by a geometric factor: $(1-\cos\theta)$ for the blue component and $\cos\theta$ for the red/purple component, where $\theta$ is the half opening angle of the blue component. Although this model neglects other important contributions such as changes in diffusion timescale, effective blackbody temperature, or angle dependence, it roughly captures a first-order correction to the assumption of spherical symmetry.  

\begin{figure*}[t]
\centering
\includegraphics[width=0.9\textwidth]{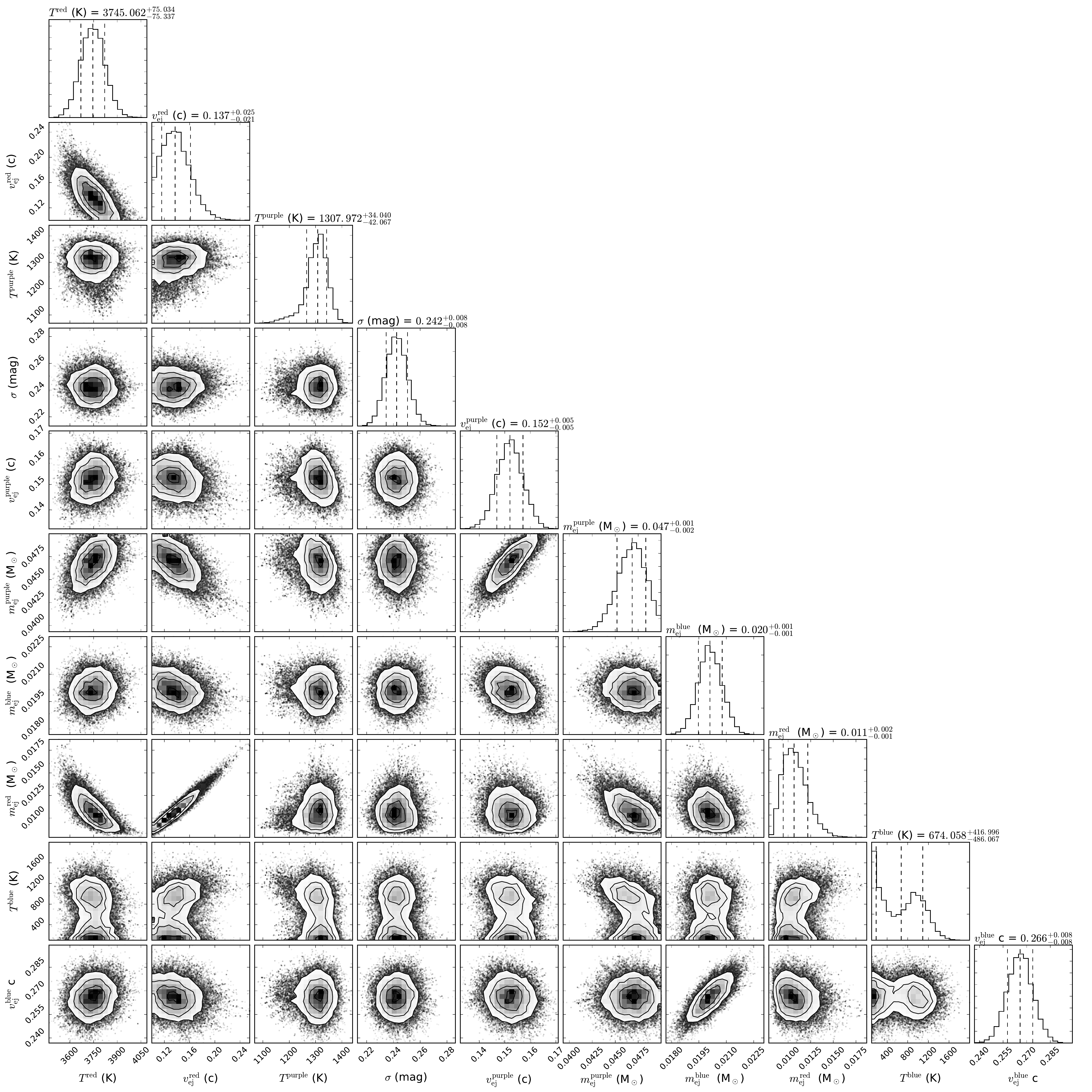}
\caption{Corner plot showing the posterior distributions of parameter realizations for the three-component model (\S\ref{sec:model}). Notable parameter degeneracies include the mass-velocity pairs of the three components, (e.g., $m_{\rm ej}^{\rm red}$ versus $v_{\rm ej}^{\rm red}$), with milder degeneracies between the temperature floors $T^{\rm red}$, $T^{\rm purple}$, and $T^{\rm blue}$ and the ejecta masses. In the former case the degeneracy is due to the ratio of the mass and velocity controlling the diffusion timescale.}\label{fig:corner}
\end{figure*}

\begin{figure*}[t]
\centering
\includegraphics[width=0.8\textwidth]{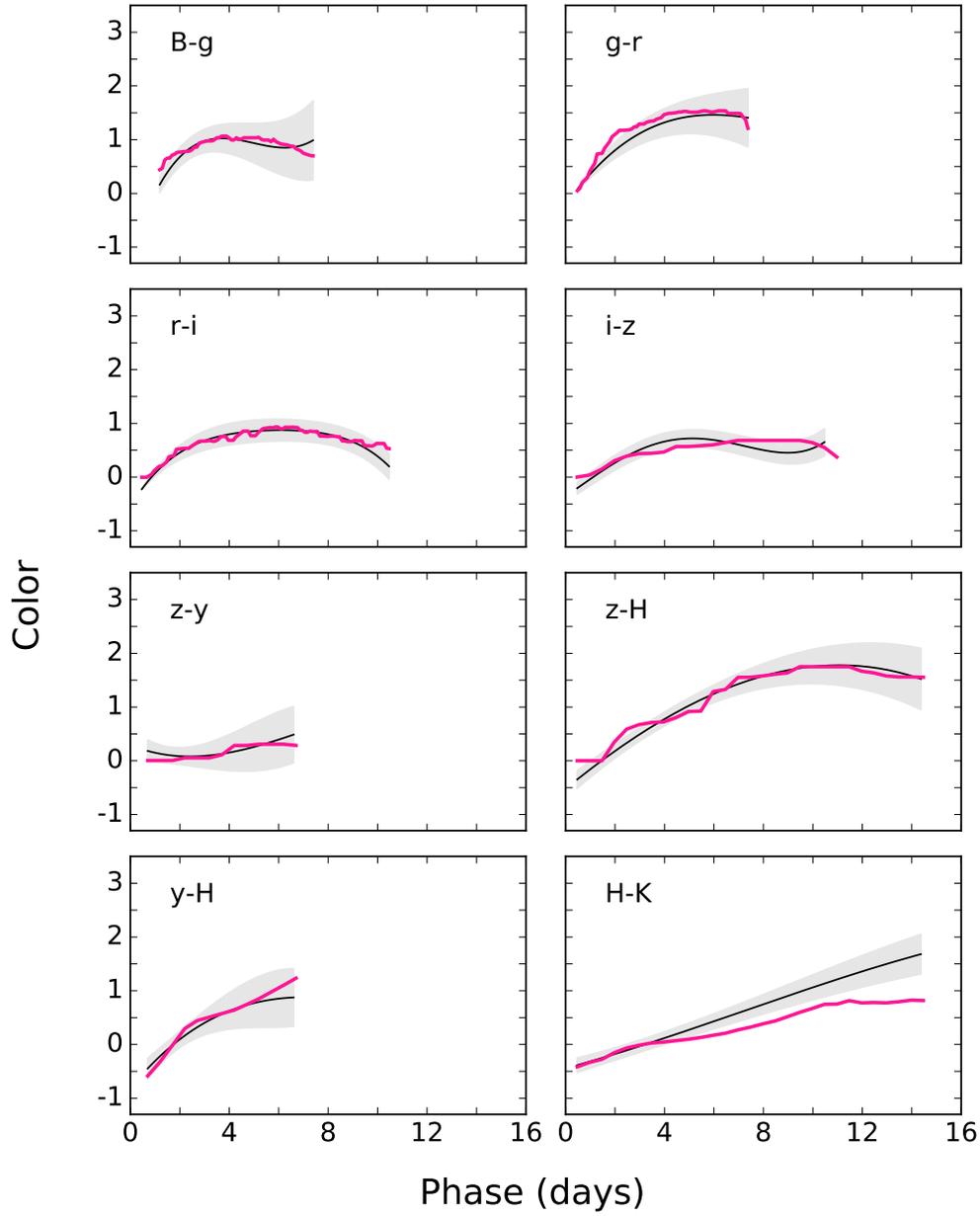}
\caption{Color evolution of the kilonova from various filter pairs. The black line shows an interpolated estimate of the observed colors, while the grey region mark the $1\sigma$ uncertainty regions, each interpolated using spline interpolation. The magenta lines are the colors for the spherically symmetric three-component model with the highest likelihood score, which have been median-filtered to minimize Monte Carlo noise.} 
\label{fig:color}
\end{figure*}

\begin{figure*}[t]
\centering
\includegraphics[width=0.95\textwidth]{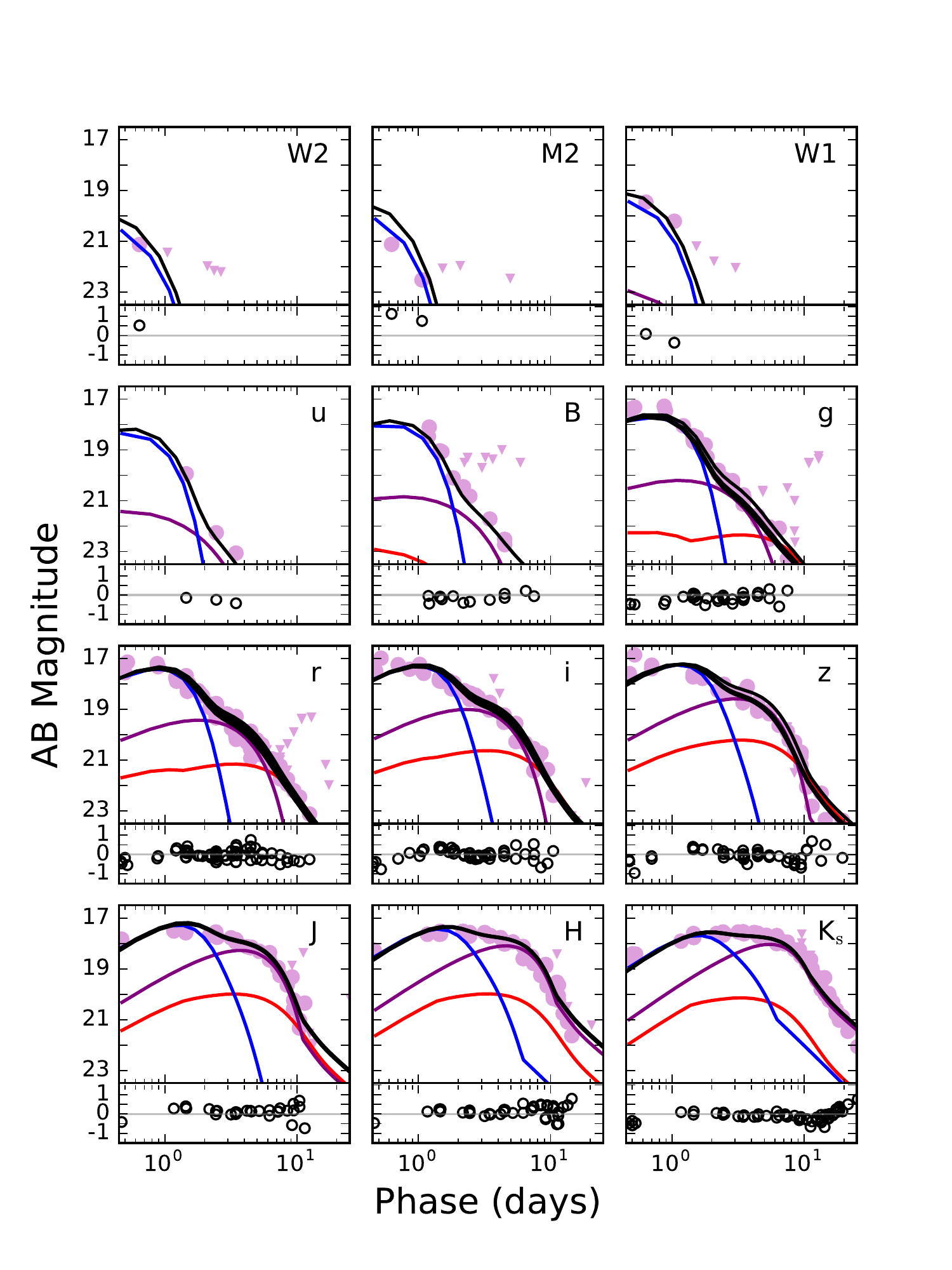}
\vspace{-0.3in}
\caption{Individual band UVOIR light curves, including the data (purple circles), the three-component best-fit model (black lines), and the individual components in the model (blue, purple, and red lines). The lower section of each panel shows the residual between the data and model. Note that some panels contain multiple black lines due to unique filter transmission functions on multiple instruments. Data originally presented in \citealt{andreoni2017,arcavi2017optical,SwopeDiscovery,cowperthwaite2017electromagnetic,diaz2017observations,drout2017light,evans2017swift,hu2017optical,kasliwal2017illuminating,lipunov2017master,pian2017spectroscopic,pozanenko2017grb170817a,ShappeeSpectra,smartt2017kilonova,tanvir2017emergence,troja2017x,utsumi2017j,valenti2017discovery}.}\label{fig:zoom}
\end{figure*}

\subsection{Fitting Procedure}

We model the combined dataset using the light curve fitting package \mosfit \citep{GuillochonMosfit,nicholl2017magnetar,villar2017}, which uses an ensemble-based Markov Chain Monte Carlo method to produce posterior predictions for the model parameters. The functional form of the log-likelihood is:
\begin{equation}
\ln\mathcal{L}=-\frac{1}{2}\sum_{i=1}^{n}\left[\frac{(O_i-M_i)^2}{\sigma_i^2+\sigma^2}-\ln(2\pi\sigma_i^2)\right]-\frac{n}{2}\ln(2\pi\sigma^2),
\end{equation}
where $O_i$, $M_i$, and $\sigma_i$, are the $i^{\rm th}$ of $n$ observed magnitudes, model magnitudes, and observed uncertainties, respectively. The variance parameter $\sigma$ is an additional scatter term, which we fit, that encompasses additional uncertainty in the models and/or data. For upper limits, we use a one-sided Gaussian penalty term.

For each component of our model there are four free parameters: ejecta mass ($M_{\rm ej}$), ejecta velocity ($v_{\rm ej}$), opacity ($\kappa$), and the temperature floor ($T_c$).  We use flat priors for the first three parameters, and a log-uniform prior for $T_c$ (which is the only parameter for which we consider several orders of magnitude). In the case of the asymmetric model, we assume a flat prior for the half opening angle ($\theta$).

For each model, we ran \mosfit for approximately 24 hours using 10 nodes on Harvard University's Odyssey computer cluster. We utilized 100 chains until they reached convergence (i.e., had a Gelman-Rubin statistic $<1.1$; \citealt{gelman1992inference}). We use the first $\simeq 80$\% of the chain as burn-in. We compare the resulting fits utilizing the Watanabe-Akaike Information Criteria (WAIC, \citealt{watanabe2010asymptotic,gelman2014understanding}), which accounts for both the likelihood score and number of fitted parameters for each model.

\begin{figure*}[t]
\centering
\includegraphics[width=0.99\textwidth]{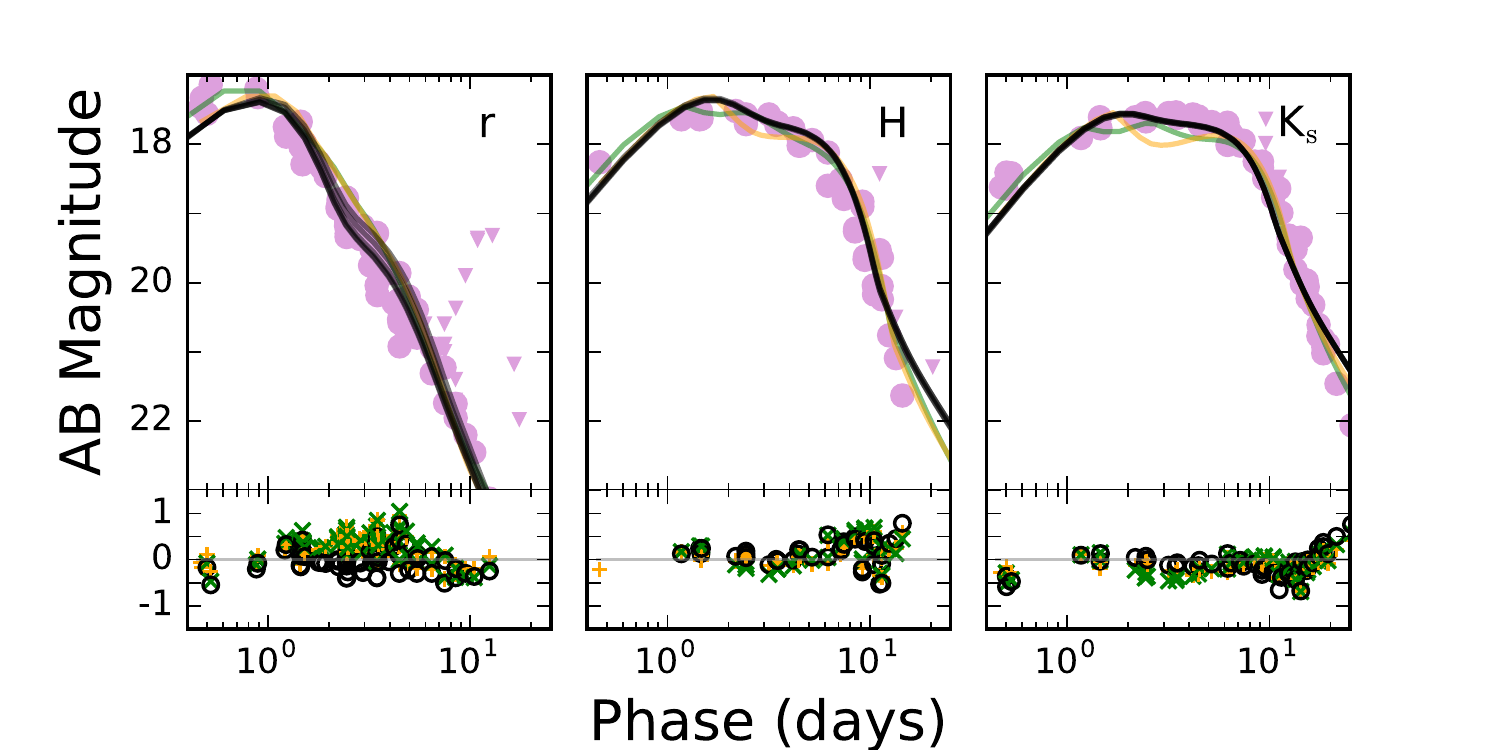}
\caption{UVOIR light curves in select bands that compare the highest likelihood model realizations of the three-component model (black lines), the two-component model (orange lines), and three-component asymmetric model (green lines).  The lower section of each panel shows the residual between the data and the three models. All models provide an overall adequate fit to the data, but the two-component predict a double-peaked structure in $K$-band that is not seen in the data.  Data originally presented in \citealt{andreoni2017,arcavi2017optical,SwopeDiscovery,cowperthwaite2017electromagnetic,diaz2017observations,drout2017light,evans2017swift,hu2017optical,kasliwal2017illuminating,lipunov2017master,pian2017spectroscopic,pozanenko2017grb170817a,ShappeeSpectra,smartt2017kilonova,tanvir2017emergence,troja2017x,utsumi2017j,valenti2017discovery}.}
\label{fig:other}
\end{figure*}

\section{Results of the Kilonova Models}
\label{sec:results}

{\setlength{\tabcolsep}{0.2em}
\begin{deluxetable*}{C{1cm}ccccccccccccccc} 
\tabletypesize{\footnotesize} 
\tablecolumns{16} 
\tablewidth{0pt} 
\tablecaption{Kilonova Model Fits \label{table:results}} 
\tablehead{ \colhead{Model} &
\colhead{M$_\mathrm{ej}^\mathrm{blue}$} & 
\colhead{v$_\mathrm{ej}^\mathrm{blue}$} &
\colhead{$\kappa_\mathrm{ej}^\mathrm{blue}$} &
\colhead{$T^\mathrm{blue}$} &
\colhead{M$_\mathrm{ej}^\mathrm{purple}$} & 
\colhead{v$_\mathrm{ej}^\mathrm{purple}$} &
\colhead{$\kappa_\mathrm{ej}^\mathrm{purple}$} &
\colhead{$T^\mathrm{purple}$} &
\colhead{M$_\mathrm{ej}^\mathrm{red}$} & 
\colhead{v$_\mathrm{ej}^\mathrm{red}$} &
\colhead{$\kappa_\mathrm{ej}^\mathrm{red}$} &
\colhead{$T^\mathrm{red}$} &
\colhead{$\sigma$} &
\colhead{$\theta$} &
\colhead{WAIC}\\ }
\startdata 
2-Comp & 0.023$^{0.005}_{0.001}$ & 0.256$^{0.005}_{0.002}$ & (0.5) & 3983$^{66}_{70}$ & - & - & - & - & $0.050^{0.001}_{0.001}$ & 0.149$^{0.001}_{0.002}$ &  3.65$^{0.09}_{0.28}$ & $1151^{45}_{72}$ & $0.256^{0.006}_{0.004}$ & & -1030\\
3-Comp & 0.020$^{0.001}_{0.001}$ & 0.266$^{0.008}_{0.008}$ & (0.5) & 674$^{486}_{417}$ & 0.047$^{0.001}_{0.002}$ & $0.152^{0.005}_{0.005}$ & (3) & 1308$^{42}_{34}$ & $0.011^{0.002}_{0.001}$ & $0.137^{0.025}_{0.021}$ & (10) & $3745^{75}_{75}$ & $0.242^{0.008}_{0.008}$ & & -1064\\
Asym. 3-Comp &  $0.009^{0.001}_{0.001}$ & $0.256^{0.009}_{0.004}$ & (0.5) & $3259^{302}_{306}$ & $0.007^{0.001}_{0.001}$ & $0.103^{0.007}_{0.004}$ & (3) & $3728^{94}_{178}$ & $0.026^{0.004}_{0.002}$ & $0.175^{0.011}_{0.008}$ & (10) & $1091^{29}_{45}$ & $0.226^{0.006}_{0.006}$ & $66^{1}_{3}$ & -1116\\
\enddata 
\end{deluxetable*}}

We fit three different models to the data: a spherical two-component model, a spherical three-component model, and an asymmetric three-component model. The results are shown in Figures~\ref{fig:lcs}--\ref{fig:other} and summarized in Table~\ref{table:results}.

For the spherical two-component model we allow the opacity of the red component to vary freely. This model has a total of 8 free parameters: two ejecta masses, velocities and temperatures, one free opacity, and one scatter term. We find best-fit values of $M_\mathrm{ej}^\mathrm{blue}=0.023^{+0.005}_{-0.001}$ M$_\odot$, $v_\mathrm{ej}^\mathrm{blue}= 0.256^{+0.005}_{-0.002}c$, $M_\mathrm{ej}^\mathrm{red}=0.050^{+0.001}_{-0.001}$ M$_\odot$, $v_\mathrm{ej}^\mathrm{red}=0.149^{+0.001}_{-0.002}c$, and $\kappa^\mathrm{red}=3.65^{+0.09}_{-0.28}$ cm$^2$ g$^{-1}$. Although the model provides an adequate fit, it predicts a double-peaked structure in the NIR light curves at $\approx 2-5$ days that is not seen in the data (Figure~\ref{fig:other}). 

Our best fitting model, the spherical three-component model, has a total of 10 free parameters: three ejecta masses, velocities and temperatures, and one scatter term. The best-fit values are $M_\mathrm{ej}^\mathrm{blue}= 0.020^{+0.001}_{-0.001}$ M$_\odot$, $v_\mathrm{ej}^\mathrm{blue}=0.266^{+0.008}_{-0.008}c$, $M_\mathrm{ej}^\mathrm{purple}=0.047^{+0.001}_{-0.002}$ M$_\odot$, $v_\mathrm{ej}^\mathrm{purple}=0.152^{+0.005}_{-0.005}c$, $M_\mathrm{ej}^\mathrm{red}=0.011^{+0.002}_{-0.001}$ M$_\odot$, and $v_\mathrm{ej}^\mathrm{red}=0.137^{+0.025}_{-0.021}c$. The parameters in this model are overall comparable to the two-component model in terms of the ejecta masses and velocities of the bluer and redder components, but here the ejecta in the redder component is distributed amongst the purple and red components. 
This model underpredicts some of the optical data at $\lesssim 1$ day and overpredicts the late time ($\gtrsim 15$ days) $K,K_\mathrm{s}$-band data; however, these deviations are less significant than for the two-component model. We additionally explored a version of this model in which the three opacities were allowed to vary freely, but found that these values fell close to our fixed values and did not significantly improve the fit.

Finally, the three-component model with an asymmetric ejecta distribution has a total of 11 free parameters: three ejecta masses, velocities and temperatures, one scatter term, and the opening angle. We find best-fit values of $M_\mathrm{ej}^\mathrm{blue}=0.009^{+0.001}_{-0.001}$ M$_\odot$, $v_\mathrm{ej}^\mathrm{blue}=0.256^{+0.009}_{-0.004}c$, $M_\mathrm{ej}^\mathrm{purple}=0.007^{+0.001}_{-0.001}$ M$_\odot$, $v_\mathrm{ej}^\mathrm{purple}=0.103^{+0.007}_{-0.004}c$, $M_\mathrm{ej}^\mathrm{red}=0.026^{+0.004}_{-0.002}$ M$_\odot$, $v_\mathrm{ej}^\mathrm{red}=0.175^{+0.011}_{-0.008}c$, and $\theta= 66^{+1}_{-3}$ degrees. This model overpredicts the intermediate time ($\approx 5$ days) optical photometry and underpredicts the early NIR photometry. Although this model has additional freedom due to the opening angle, the ejecta masses become linked through this additional parameter. Due to the simplicity of the asymmetric model, we do not take the derived parameters and uncertainties at face value, and instead use them as a guide for the effects of asymmetry. We find that an asymmetric ejecta distribution leads to masses that are $\approx 50\%$ lower than in the spherical case. 

We note that the inferred value of $\theta$ is consistent with the blue component being visible at an orbital inclination angle of $\approx 20-50^\circ$, as inferred from a comparison of the GW waveform to the source distance, and from an analysis of the radio and X-ray data in the context of an off-axis jet \citep{PhysRevLett.119.161101,alexander2017,guidorzi2017,hallinan2017,margutti2017,murguiaberthier2017}. The relatively large angle is also consistent with the low polarization found by \cite{covino2017}. 

Our spherical three-component model realization of highest likelihood (the ``best fit'') is shown with the complete dataset in Figure~\ref{fig:lcs}, and its corresponding corner plot is shown in Figure~\ref{fig:corner}.  Overall the model provides a good fit to the complete dataset.  We find that most parameters are constrained to within $\lesssim 10$\%. The true errors in our models are likely larger, suggesting that the uncertainty is likely dominated by systematic effects (e.g., uncertainty in thermalization efficiency, heating rate, etc.).

We show the individual filters with each of the three components (and their sum) in Figure~\ref{fig:zoom}. We find that the blue component dominates across all bands at $\lesssim 2-3$ days, while the purple component dominates at later times. Because of its low ejecta mass, the reddest component is sub-dominant at all times but contributes necessary flux to the redder bands at late times.

We explore the color evolution of our model compared to that of the kilonova in Figure~\ref{fig:color}, and again find that the model largely recovers the rapid color evolution, although it slightly deviates from the observed NIR colors at $\gtrsim 12$ days.  Finally, we show specific representative filters ($r$, $H$, $K_\mathrm{s}$) with a comparison of all three models in Figure~\ref{fig:other}. Although the differences are subtle, the three-component model provides a statistically better fit to the overall light curves.  We stress that the overall success of all three models is remarkable given the extensive scope of the data in time and wavelengths, and the simplifying assumptions in our analytic approach.

\section{Discussion and Implications}
\label{sec:implications}

Our best fit three-component model, dominated by an intermediate purple component, is consistent with previous findings (e.g., \citealt{cowperthwaite2017electromagnetic,nicholl2017electromagnetic,chornock2017electromagnetic}).  Compared to our previous modeling presented in \citet{cowperthwaite2017electromagnetic}, both the blue and purple ejecta masses and the purple velocity increased by $\approx40$\%. The other parameters remained within $\approx 1\sigma$ of the previously reported values. The uncertainties on the fitted parameters have decreased by $\approx 10-50$\% due to the dramatic increase in the number of data points. Our inferred total ejecta mass of $\approx 0.078$ M$_\odot$, somewhat higher than the values inferred by several groups based on their individual subsets of the dataset we modeled here ($\approx 0.02-0.06$ M$_\odot$; \citealt{kasliwal2017illuminating,2017arXiv171005434K,tanaka2017kilo}). Additionally, modeling of the optical and NIR spectra indicates that the early blue emission is best described by material with a gradient of lanthanide fraction, with the fraction increasing with time \citep{nicholl2017electromagnetic,chornock2017electromagnetic}. This is consistent with our findings that the purple component begins to dominate the UVOIR light curves at $\approx 2-3$ days post-merger. 

The inferred high velocity of the blue ejecta is most naturally explained by relatively proton-rich (high electron fraction, $Y_{e}$) polar dynamical ejecta created by the shock from the collision between the merging neutron stars (e.g., \citealt{Oechslin+06,Bauswein+13,Sekiguchi+16,Radice+16}).  In this scenario, the inferred high ejecta mass ($\approx 0.02$ M$_{\odot}$) is indicative of a small neutron star radius of $\lesssim 12$ km when compared to the results of numerical simulations (\citealt{Hotokezaka+13,Bauswein+13}; see also \citealt{nicholl2017electromagnetic}).  Alternatively, the blue ejecta could arise from a neutrino-heated outflow from a hyper-massive neutron star (e.g., \citealt{rosswog2002,Dessart+09}), although the high mass and velocity of the blue ejecta greatly exceed the expectations from a standard neutrino wind and would likely require additional acceleration of the wind by strong magnetic fields (e.g., \citealt{Metzger+08Ye}).

The red ejecta component could in principle originate from the dynamically-ejected tidal tails in the equatorial plane of the binary (e.g., \citealt{rosswog1999,Hotokezaka+13}), in which case the high ejecta mass would require a highly asymmetric merger  with a binary mass ratio of $q\lesssim 0.8$ \citep{Hotokezaka+13}.  However, the velocity of this component ($\approx 0.1 c$) is much lower than those typically found in simulations of NS mergers with extreme mass ratios ($\approx 0.2-0.3 c$; \citealt{Kilpatrick+17}) potentially disfavoring this explanation. Additionally, our large mass estimate is on the upper end of the dynamical ejecta mass estimated by \citet{2017arXiv171005836T}, suggesting that not all of this mass is dynamically ejected.

A more promising source for the red and purple ejecta components is a delayed outflow from the accretion disk formed in the merger \citep{Metzger+09,Fernandez&Metzger13,Perego+14,Just+15,Siegel&Metzger17}, for which the outflow velocity is expected to be $\approx 0.03-0.1 c$.  The relatively high neutron abundance of this matter ($Y_{e}\lesssim 0.25-0.3$ as needed to synthesize lanthanide nuclei) would be consistent with the moderate amount of neutrino irradiation of the outflow from a black hole accretion disk \citep{Just+15} but would disfavor a particularly long-lived ($\gtrsim 100$ ms) hyper-massive or supra-massive neutron star remnant (\citealt{Metzger&Fernandez14,murguia2014,Kasen+15,Lippuner+17}; see also \citealt{Margalit&Metzger17}).  In this context, the properties of the red/purple ejecta provide evidence for a relatively prompt formation of a black hole remnant.   

The asymmetric model indicates a half-opening angle for the blue component of $\theta\approx 66^\circ$.  This is consistent with the blue component being visible given the inclination angle of the system inferred both from a comparison of the GW waveform and the distance of the event, and from off-axis jet models of the radio and X-ray light curves ($\approx 20-50^\circ$; \citealt{PhysRevLett.119.161101,alexander2017,margutti2017}). Our simple asymmetric model suggests that the total ejecta mass may be $\approx 50\%$ smaller than inferred in the spherical model. The effects of other simplifying assumptions, such as the blackbody SED and constant opacities as a function of time and wavelength, should be explored in future work.

Finally, we compare our inferred total ejecta mass to the amount necessary to reproduce the Milk Way $r$-process production rate using the updated BNS merger rate inferred from Advanced LIGO of $R_0=1500^{+3200}_{-1220}$ Gpc$^{-3}$ yr$^{-1}$ \citep{PhysRevLett.119.161101} following a similar methodology as \citet{cowperthwaite2017electromagnetic} and \citet{kasen2017}. For light $r$-process nuclei, the primary source of ejecta in our three component model, the inferred Milky Way production rate is $\dot{M}_\mathrm{rp,A\lesssim 140}\approx 7\times10^{-7}$ M$_\odot$ yr$^{-1}$ \citep{qian2000}. Combining this with the BNS rate and density of Milky Way-like galaxies ($\approx 0.01$ Mpc$^{-3}$), we estimate the Milky Way rate of BNS mergers as $R_\mathrm{MW}\approx 150$ Myr$^{-1}$. Thus, the average ejecta mass necessary for a blue/purple kilonova is $\dot{M}_\mathrm{rp,A\lesssim140}/R_\mathrm{MW}\approx 5\times 10^{-3}$ M$_{\odot}$, with an uncertainty of about a factor of $\approx 5$ due to the large range of $R_0$.  For heavy $r$-process elements (our red component), the Milky Way inferred production rate is $\dot{M}_\mathrm{rp;A\gtrsim140}\approx 10^{-7}$ M$_\odot$ yr$^{-1}$ \citep{bauswein2014}. The average ejecta mass necessary for a red kilonova is therefore $\dot{M}_\mathrm{rp,A\lesssim140}/R_\mathrm{MW}\approx 7\times10^{-4}$ M$_{\odot}$, again with an uncertainty of about a factor of 5. In both cases, this order of magnitude estimate is about a factor 10 times smaller than our estimated ejecta masses for this event, although the rate errors (and potentially lower ejecta masses in the asymmetric case) are large enough to account for the discrepancy\footnote{Our results are consist with those found in \citealt{2017arXiv171005836T}.}. However, we note that the ratio of red to blue/purple ejecta masses in our model, $\approx 0.16$, is in good agreement with the relative production rates of $A\gtrsim 140$ and $A\lesssim 140$ nuclei in the Milky Way.

If the BNS merger rate from future events is shown to be at the high end of the current estimates, the results inferred here would indicate that a large fraction of synthesized $r$-process material may remain in the gas phase within the ISM or escape the galaxy entirely via galactic winds \citep{shen2015}. It may also suggest that the kilonova in GW170817 is an outlier in terms of total $r$-process material produced. Future events will clarify the population parameters of kilonovae.

\section{Conclusions}

We presented the first effort to aggregate, homogenize, and uniformly model the complete UV, optical and NIR dataset for the electromagnetic counterpart of the binary neutron star merger GW170817, allowing us to better determine the likely combinations of parameters responsible for the observed kilonova. We are able to remove systematic offsets from several datasets and to identify outlying data points, providing the community with cleaned and uniform photometry for future analyses. Our key findings are as follows:

\begin{itemize}
  \item We present 647 photometric measurements from the kilonova accompanying the binary neutron star merger GW170817, spanning from 0.45 to 29.4 days post-merger and providing nearly complete color coverage at all times. We make the homogenized dataset available to the public in Table~\ref{table:data}, in the OKC, and through \url{https://kilonova.org/} 
  
  \item The kilonova UVOIR light curves are well fit by a spherically symmetric, three-component model with an overall ejecta mass of $\approx 0.078$ M$_\odot$, dominated by light $r$-process material ($A<140$) with moderate velocities of $\approx 0.15 c$. 
  
  \item We find evidence for a lanthanide-free component with mass and velocity of $\approx 0.020$ M$_\odot$ and $\approx 0.27c$, respectively. This component is indicative of polar dynamical ejecta, and hence a BNS origin (instead of NS-BH). The large ejecta mass implies a small neutron star radius of $\lesssim 12$ km.
  
  \item The mass and velocities of the purple/red components are consistent with a delayed outflow from an accretion disk formed in the merger. This disfavors a long-lived ($\gtrsim 100$ ms) hyper-massive neutron star remnant and provides evidence for relatively prompt formation of a black hole remnant.
  
  \item The asymmetric model extension implies that the total ejecta mass may be up to a factor of 2 times lower than for the symmetric model.
  
  \item Given the large uncertainties in BNS merger rates, we find that the $r$-process production rates are comfortably above the Galactic production rate, consistent with the idea that BNS mergers are the dominant source of $r$-process nucleosynthesis in the universe.
\end{itemize}

The sheer size of the dataset for this event, which was the subject of unprecedented follow-up efforts by the observational astronomy community, represents a departure from typical transient events, allowing for more detailed modeling than typically feasible.  Although future observing runs of Advanced LIGO/Virgo will lead to many more kilonova detections, it is likely that this event will remain one of the best-observed objects for years to come due to its vicinity and hence ease of follow-up.  Thus, the broad UVOIR dataset collected by multiple teams, and aggregated and homogenized here, will be an invaluable resource to explore questions about kilonova phenomenology that may be otherwise intractable using more sparsely sampled data.

\acknowledgments
We thank the anonymous referee and the larger community for valuable feedback on this work. The Berger Time-Domain Group at Harvard is supported in part by the NSF through grant AST-1714498, and by NASA through grants NNX15AE50G and NNX16AC22G. VAV acknowledges support by the National Science Foundation through a Graduate Research Fellowship. This research has made use of NASA's Astrophysics Data System.

\software{%
\texttt{astrocats} \citep{OSCRef}, 
\texttt{matplotlib} \citep{matplotlib}, 
\mosfit \citep{GuillochonMosfit};
\texttt{numpy} \citep{numpy}, 
\texttt{scipy} \citep{jones_scipy_2001}
}

\clearpage 
\startlongtable
\begin{deluxetable*}{c c c c c C{1cm} C{1cm} c C{2.7cm} c}
\tabletypesize{\footnotesize} 
\tablecolumns{10} 
\tablewidth{0pt} 
\tablecaption{Photometric Data \label{table:data}} 
\tablecomments{We request that any use of the data in this table includes appropriate citation to the original papers, as well as to our compilation.}
\tablenotetext{a}{New magnitude value used in modeling.}
\tablenotetext{b}{Difference between new value and originally reported value.}
\tablenotetext{c}{Photometry listed with an ``x'' is not included in our model fit,  photometry listed with an ``o'' has been visually flagged as an outlier, photometry reported in multiple sources with unique reduction routines are listed with an ```r'', photometry generated by averaging repeated photometry is listed with an ``a'', and photometry used in modeling is listed with an ``*''. }
\tablehead{ \colhead{MJD} &
\colhead{Phase} & 
\colhead{Instrument} &
\colhead{Telescope} &
\colhead{Filter} &
\colhead{AB Mag\tablenotemark{a}} &
\colhead{1$\sigma$ Err} &
\colhead{$\Delta$(Mag)\tablenotemark{b}} &
\colhead{Ref.}  &
\colhead{Note\tablenotemark{c}}\\}
\startdata 
57982.981 & 0.452 &  E2V 4kx4k ccd & Swope & i & 17.48 & 0.02& 0 & \citeauthor{SwopeDiscovery}&*\\
57982.990 & 0.461 & FourStar & Magellan & H & 18.26 & 0.15 &  0 &\citeauthor{drout2017light}&*\\
57982.993 & 0.464 &  Alta U47+ & Prompt5 & r & 17.46 & 0.03 &  0 &\citeauthor{valenti2017discovery}&*\\
57982.999 & 0.470 & VIRCAM & VISTA & Ks & 18.62 & 0.05 & 0 & \citeauthor{tanvir2017emergence}&*\\
57983.000 & 0.471 & FourStar & Magellan & J & 17.83 & 0.15 & 0 & \citeauthor{drout2017light}&*\\
57983.000 & 0.471 & LDSS & Magellan & V & 17.35 & 0.02 & 0 & \citeauthor{ShappeeSpectra}&*\\
57983.000 & 0.471 & LDSS & Magellan & r & 17.33 & 0.02 & 0 & \citeauthor{ShappeeSpectra}&*\\
57983.000 & 0.471 & LDSS & Magellan & z & 17.67 & 0.03 & 0 & \citeauthor{drout2017light}&*\\
\enddata 
\end{deluxetable*}

\bibliographystyle{yahapj}
\bibliography{new.ms}

\end{document}